\documentclass[final,5p,times,twocolumn]{elsarticle}

\usepackage{graphicx}
\usepackage{amssymb}

\usepackage{lineno}

\usepackage{amsmath}

\usepackage{esvect}
\usepackage{textcomp}

\usepackage{float}

\usepackage[ruled, linesnumbered,noend]{algorithm2e}
\usepackage[dvipsnames]{xcolor}
\usepackage[utf8]{inputenc}
\usepackage[T1]{fontenc}
\usepackage{latexsym}
\usepackage{amssymb}
\usepackage{listings}
\usepackage{caption}

\usepackage[american]{babel}
\usepackage{csquotes}

\usepackage{cleveref}[2012/02/15]
\crefformat{footnote}{#2\footnotemark[#1]#3}

\usepackage{xpatch}

\xpatchbibmacro{name:andothers}{%
  \bibstring{andothers}%
}{%
  \bibstring[\emph]{andothers}%
}{}{}

\newcounter{examplecounter}

\newenvironment{example}
{        
    \vspace{3pt}
    \refstepcounter{examplecounter}%
  \textbf{Example \arabic{examplecounter}}%
  \quad
}{
\vspace{3pt}
}

\usepackage{amssymb}
\usepackage{amsmath}
\usepackage[hang,flushmargin]{footmisc}
\usepackage{caption}

\newtheorem{definition}{Definition}

\makeatletter
\newcommand{\removelatexerror}{\let\@latex@error\@gobble}
\makeatother

\makeatletter
\newcommand*{\rom}[1]{\expandafter\@slowromancap\romannumeral #1@}
\makeatother

\lstdefinelanguage{code}{
  morekeywords={let,in,def,aspect,before,after,pointcut,public,privileged,protected,declare,parents, call,target,implements,throw,new,for,class,forAll,exists,Boolean,return,break,executeCaller,true,
  false,pre,if,then,else,endif,String,join,select,from,where,with,create,temporary,table,and,
  or,as,on,case,when,end,union,iterate,intersection,symmetricDifference,self,includes,function, Text,Set,boolean, show,insert, into,div,mod,Integer, by, having, on, not, var, while, continue,switch,execute,abort,delete,eval,this,input, output,process},
  basicstyle=\footnotesize\usefont{T1}{pcr}{m}{n}\selectfont,
  keywordstyle=\footnotesize\usefont{T1}{pcr}{b}{n}\selectfont,
  identifierstyle=\footnotesize\usefont{T1}{pcr}{m}{n}\selectfont,
  commentstyle=,
  stringstyle=\footnotesize\usefont{T1}{pcr}{m}{n}\selectfont,
  numberstyle=\footnotesize,
  tabsize=2,
  frame=lines,
  upquote=true,
  xleftmargin=10pt
}

\journal{Computer Networks}

\begin{document}

\begin{frontmatter}

\title{Decentralizing Privacy Enforcement for Internet of Things Smart Objects}

\author{Gokhan~Sagirlar,Barbara~Carminati,and~Elena~Ferrari}

\address{DISTA, University of Insubria, Italy.}
\cortext[cor1]{
E-mail addresses: \\
Gokhan Sagirlar: gsagirlar@uninsubria.it\\
Barbara Carminati: barbara.carminati@uninsubria.it\\
Elena Ferrari: elena.ferrari@uninsubria.it}



\begin{abstract}
Internet of Things (IoT) is now evolving into a loosely coupled, decentralized system of cooperating smart objects, where high-speed data processing, analytics and shorter response times are becoming more necessary than ever. Such  decentralization has a great impact on the way personal information generated and consumed by smart objects should be protected, because, without centralized data management, it is more difficult to control how data are combined and used by smart objects. To cope with this issue, in this paper, we propose a framework where users of smart objects can specify their \textit{privacy preferences}. Compliance check of user individual privacy preferences is performed directly by smart objects. Moreover, acknowledging that embedding the enforcement mechanism into smart objects  implies some overhead, we have extensively tested the proposed framework on different scenarios, and the obtained results show the feasibility of our approach.
\end{abstract}


\end{frontmatter}


\section{Introduction}\label{sec:introduction} 

Internet of Things (IoT) technologies are revolutionizing our daily lives \cite{iotVision}, building around us a pervasive environment  of  {\em smart objects}  able, not only to sense data, but also to interact with other objects and  to  aggregate data sensed through different sensors. This allows smart objects to locally create new knowledge, that could be used to  make decisions, such as   quickly trigger actions on  environments, if needed. Smart objects are very heterogeneous in terms of data sensing and data processing capabilities. 
Some of them can only sense  data, others  can perform basic or complex operations on them. Such  a scenario enacts the transition from the Internet of Things to the Internet of Everything, a new definition of IoT seen as a loosely coupled, decentralized system of cooperating smart objects, which leverages on alternative architectural patterns with regards to the centralized cloud-based one, such as fog computing. Such a trend towards decentralization reduces the amount of data that is transferred to the cloud for processing and analysis, and can also be instrumental to  improve security and privacy of the managed data, a major concern in the IoT scenario. However, decentralization, if not properly governed,  might also imply loss of control over the data, with consequences on individual privacy.

In this paper, we focus on the challenging issue of designing a  decentralized  privacy enforcement mechanism, where compliance check of user individual privacy preferences is performed directly by smart objects, rather than by a central entity. Restrictions on devices' capabilities  let us discard existing proposals for decentralized access control (e.g., \cite{skarmeta,Sicari2016133,authACIot,distributedACPrivacy}), as these heavily rely on cryptographic primitives. 
Previously, in \cite{ourPaper}, we addressed the problem of specifying and enforcing privacy preferences in the IoT scenario, but for a centralized architecture, that is, a scenario where devices have only the capability to sense data and send them to a data center for being analyzed. In this setting, the enforcement monitor analyzes every consumer query and decides if the privacy policy of the consumer satisfies the privacy preferences specified by owners of devices generating the data.
Compared to this approach, decentralized privacy enforcement scenario requires to address several new important research challenges, as  smart objects are characterized by heterogeneous processing capabilities. 
To address these challenges, in this paper, we extend the privacy preference model proposed in \cite{ourPaper}, by designing a set of privacy meta-data that are used by smart objects for locally checking privacy preferences and  for locally enforcing user privacy preferences at smart object level.  
Smart objects are thus able to  derive privacy meta-data for newly created data items, keep track of the operations performed over data items,  denoted as {\em history}, in order to ease privacy preference enforcement, and, finally, check compliance of the privacy policy of the data consumer with the privacy preferences of data items. To the best of our knowledge this is the first work  proposing a decentralized enforcement of privacy preferences able to work  locally at smart object level.

Acknowledging that embedding the enforcement mechanism into smart objects might imply some overhead, we have extensively tested the proposed framework. In doing the experiments, we have considered several scenarios, by  varying  the complexity of the privacy preferences,  smart object networks, and  evaluated queries. The experiments allow us to asses the feasibility of the proposed approach in a variety of application domains. 

The remainder of this paper is organized as follows. 
Section \ref{sec:system} describes the system model and design assumptions of the proposed privacy preserving framework.
Section \ref{sec:privacyModel} introduces the privacy preference model. Section \ref{sec:enforcement} presents the proposed enforcement mechanism. Experimental results are illustrated in Section \ref{sec:experiments}. 
Section \ref{sec:related} discusses related work, whereas Section \ref{sec:conclusion} concludes the paper.

\section{System Model and Assumptions}\label{sec:system}
The main purpose of the framework is to enhance  the  privacy of IoT users making them able to set and enforce their privacy preferences. The privacy enforcement mechanism relies on some privacy meta-data (encoding user privacy preferences, data categories, and data history), generated by smart objects. 
Data generated by smart objects are modelled as data streams. 
Moreover, we assume that smart objects are able to perform  a set of   SQL-like operations (i.e., projection ($\Pi$), aggregation ($\Sigma$), join ($\Join$) and selection ($\sigma$)) over data streams, as SQL represents the query model adopted by many data processing systems in the IoT domain \cite{iotSql}. 
These SQL operators are continuously  evaluated on every incoming  tuple, whereas operations like join and aggregation are processed  on groups of tuples organized as sliding windows. 
In the following, we present the system model and our design assumptions.

\subsection{System model}
We consider a decentralized IoT system consisting of the following entities:

\textbf{Subjects.}
We consider two types of subjects, namely  data owners and data consumers.
Data owner is the user that owns an IoT smart object.
We assume that data owners define individual privacy preferences over their data (cfr.  Section \ref{sec:privacyModel} for a detailed discussion) in order to control how their data are distributed and processed. In contrast, data consumer  consume raw or processed data from IoT systems.
We assume that data consumers adopt privacy practices that specify how the consumer will use the users' data.

\textbf{Smart Objects.} 
Smart objects are able to sense, process and aggregate data, and interact with other smart objects.
However, they are extremely heterogeneous regarding their capabilities and roles in  the IoT ecosystem and this also impacts the privacy checks they can perform.
In particular, we need to classify smart objects based on their capabilities so as to assign to each of them a possible $role$, and corresponding  function in  the proposed decentralized privacy preference enforcement mechanism. 
On the basis of several standardization recommendations from different organizations  \cite{Nistir, ituStandard}, we propose a smart objects taxonomy based on the object \textit{smartness}, aka abilities in sensing and processing individual data, as follows: 

-- {\em First Level of Smartness:}\label{level1} 
at this level, devices have  a very limited capacity, being able only to act as  basic sensors, for sensing  data from the environment.   

-- {\em Second Level of Smartness:}\label{level2}
at the second level, devices have a limited capacity and computing power, making them only able to  perform basic operations by their own, like projection and selection, which do not require window-based processes that are working on tuples organized as sliding windows, such as join or aggregation.
These devices might  be virtual or physically manufactured, e.g., a field-programmable gate array (FPGA) or hard-coded  devices.  

-- {\em Third Level of Smartness:}\label{level3}
at this level,  we group those devices that have enough capacity and computing power for performing  complex operations by their own, including  window-based operations.    

According to this taxonomy, a  {\em smart object (SO)} may play three different roles in the proposed decentralized privacy enforcement:\footnote{We postpone the reader to Section  \ref{sec:enforcement} for details about processes and algorithms implemented for each role.}

-- {\em Sensing Smart Objects:} 
    the main objective of this type of  smart objects is sensing data from the environment. Typically,  a  smart object with a first level of smartness will play this role. 
        
-- {\em Processor Smart Objects:}
    this type of  smart objects aim at  creating new information from sensed data by performing operations over them. Complexity of these operations  depends on the smartness level of the object. As such, a  smart object with second or third level of smartness will be able to play this role. 
    
-- {\em Consumer Smart Objects:}
    a smart object playing this role is in charge of verifying user privacy preferences before sending data to consumers.  This process is called privacy preference compliance (see Section \ref{complianceAnalysis} for more details).  
Such role can be taken by SOs with third level of smartness (e.g., gateway devices, such as Dell Edge Gateways or HPE Edgeline Gateways).

\begin{example}\label{exp:smartObjects}
Throughout the paper, we will use as an example  a smart home system. We assume that the smart home is equipped with a smart  heating and  electricity usage control systems. The smart heating system is built on top of a network of smart objects  able to sense temperature values from several rooms, adjust temperature, if needed, and share data with the heating  company. The smart electricity usage system relies on a  network of smart objects  able to sense electricity usage data, locally compute  usage trends, and  send them  to the electricity distribution company  (see Figure \ref{fig:smarthome}).  In this  scenario, sensing smart objects are those devices used for measuring electricity usage and sensing room temperature, whereas processor smart objects are those used to locally aggregate data sensed by sensing SOs, with the purpose of generating trends of electricity usage as well as average rooms temperature.  Finally, smart objects sending collected information to electricity and heating system companies can be categorized  as consumer smart objects, as their main role is to send data to consumers. 

 \begin{figure} [h] 
	\centering
	\includegraphics[width=0.5\textwidth]{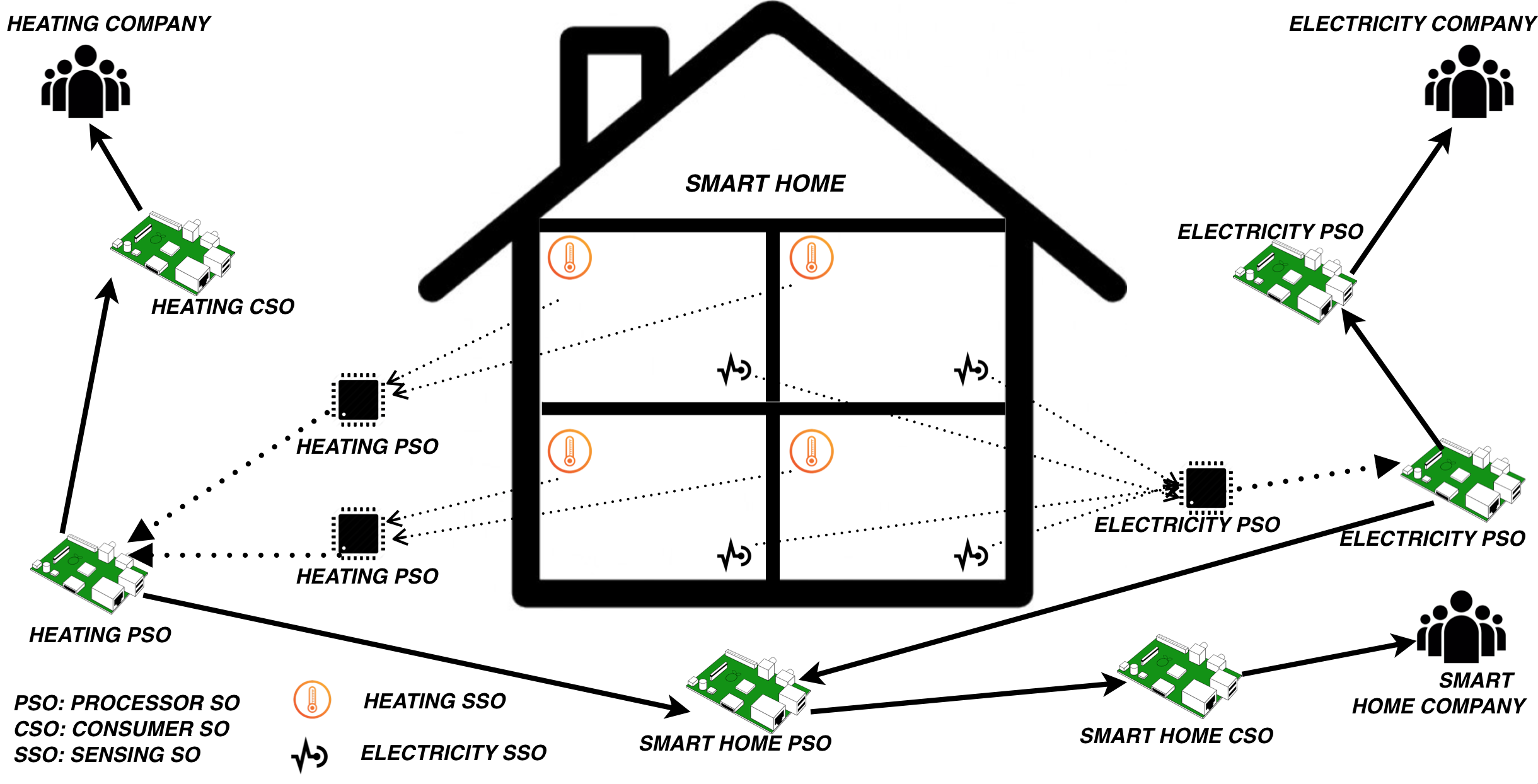}
	\caption{Smart home scenario} \label{fig:smarthome}
\end{figure}

\end{example}

\subsection{Security assumptions}\label{sec:assumptions}
Security vulnerabilities, if not properly addressed and prevented, may allow adversaries to damage correct execution of IoT systems and the proposed privacy preserving mechanism.
In designing the privacy preserving mechanism, we have assumed that smart objects conduct a set of defence strategies to protect the IoT system from security threats and attacks.
Let us first define the adversary model.
We assume that an adversary is capable to block and intercept communication, and is also able to overhear, inject and delete messages exchanged between smart objects, as well as analyze communication patterns.
Moreover, we assume that sensing  and processor SOs are susceptible to be compromised by adversaries and thus can act maliciously, whereas consumer SOs are  honest, i.e., they  behave as expected according to the proposed algorithms. 
This assumption is based on the conjecture that IoT gateway devices, such as Dell Edge Gateways\footnote{$dell.com/us/business/p/edge-gateway$}, or HPE Edgeline Gateways\footnote{$hpe.com/emea_europe/en/servers/edgeline-iot-systems$} hosting consumer SOs are securely produced by their manufacturers and have high resources.  Indeed, as stated by \cite{securityAnalysisIoT}, a vulnerability in the hardware of such devices will inevitably  threat the whole system. 
Therefore, we assume that device manufactures will do their best in designing secure gateway devices.
Given the adversary model, the privacy preserving mechanism relies on the following defense techniques that are assumed to be conducted by smart objects:

\textbf{Network monitoring and intrusion detection systems.} 
We assume that consumer SOs are complemented with network monitoring (e.g., \cite{networkBehAnalysisP2P}) and intrusion-detection mechanisms (e.g., \cite{intrusionIoT}).
Because such mechanisms help to detect and reduce effects of \textit{Denial of Service (DoS)} and \textit{node compromise} attacks.
Indeed, as our approach relies on machine to machine communication among smart objects in which the operations of one smart object, e.g., meta-data generation and update, mostly depend on the data sent by another smart object, DoS attacks may affect the correct execution of the protocols.
Furthermore, compromised and malicious processor SOs can alter the derivation of  new privacy meta-data with the result of not correctly enforcing privacy requirements of the data owners. 
By virtue of equipping consumer SOs with intrusion detection mechanisms, we assume that consumer SOs analyze the network and detect compromised and infected devices acting maliciously or violating policies. 


\textbf{Lightweight encryption and authentication mechanisms.} 
We also  assume that all smart objects are equipped with encryption mechanisms (such as \cite{lightweightEncryption,cipherSecCertf}) that make use of lightweight cryptographic primitives (e.g., hash, nonce, etc), and thus able to encrypt the data and the meta-data before sending them.
In such a way, that data cannot be disclosed and privacy meta-data cannot be compromised during the communication.

We also assume that smart objects are equipped with hop by hop authentication protocols, such as \cite{heer2008alpha}. 
This is a a way to contrast  \textit{impersonation} attacks, where an adversary might aim to seize the identity of a legitimate smart object, such as access credentials of the device, to act on behalf of the legitimate device (e.g. injecting fake data) that may compromising privacy preference enforcement.

\textbf{Clone detection protocols.} 
We assume that smart objects are equipped with a globally aware distributed clone detection protocol (such as  \cite{cloneAttacksDist}).
Cloned smart objects may generate fake data and affect the correct execution of the privacy protection protocols, as well as they  can be used to disable and compromise honest smart objects and thus damage privacy enforcement.


\section{The Privacy Model}\label{sec:privacyModel}
In this section, we present the privacy preference model and an approach to automatically generate privacy preferences for new generated data.

\subsection{Privacy Preferences}
Privacy preferences are specified according to the model proposed in \cite{ourPaper}, which has been designed to cope with two relevant  challenges that IoT ecosystems pose on the management of individual privacy. 

The first is that, due to the complexity of the data flows among IoT devices, users easily lose the control on how their data are distributed and processed. This is even  made worse considered that data generated by different IoT devices can be fused  to infer new information on individuals. To give individuals more control, we support privacy preferences
that are able to specify additional constraints with regards to those  that can be expressed by standard privacy languages. 
In particular, our privacy preferences  make a user able to pose conditions not only  on which portion can be collected, for which  usage purpose, for how long, and by whom  (like traditional privacy policies and preferences), but also make him/her able to pose limitations on how  data can be used to derive new information (e.g.,  by explicitly stating  what cannot be derived from users data by analytics processes). 

The second issue the privacy model considers is related to the management of new derived data. Indeed, although  privacy preferences help individuals at regulating the derivation of
new data, they do not regulate how these new data will be managed. At this purpose, the privacy model implements a strategy for the automatic definition of additional privacy preferences for newly derived data. The new preference is defined by taking into account the privacy preferences associated  with each single piece of data used for computing the new derived data.

Similar to other privacy preference models, we organize  both purposes and data categories into trees. This simplifies the specification of privacy preferences, in that a preference stated for a node in the tree representing a purpose/category  (e.g., the physical state data category) implicitly holds also for  all nodes in its subtree (e.g., heart beat, blood pressure, etc.). Moreover, besides making  individual able to specify the purposes for which is allowed the data usage, the model aims at making him also able to pose limitation. As such, the  privacy preference allows the specification of those purposes for which the access is explicitly prohibited.  
This is represented via an element, called {\em intended purposes -- ip}, which is modeled as a pair (\textit{Aip}, \textit{Exc}), with the following semantics: all requests with a purpose that  descends from  elements in \textit{Aip}  are authorized, except for those that descend from any element in \textit{Exc} in the purpose tree.

Besides intended purposes, privacy preferences allow users to explicitly state which portions of their personal data can be  combined/aggregated with other data. Such requirements are expressed in terms of allowed/denied data categories, modelled through the so called {\em joint access constraints -- jac}. More formally, a joint access constraint  for an attribute \textit{a}  is a tuple $\langle$\textit{Adc, Exc, ip}$\rangle$, where \textit{Adc} is the set of  data categories that can be jointly accessed with \textit{a},
 \textit{Exc} is a set of data categories that specialize elements in \textit{Adc} and that cannot be jointly accessed, whereas \textit{ip} is an intended purpose. This constraint specifies that the categories that descend from those specified in \textit{Adc} can be jointly accessed with attribute \textit{a} for the purposes authorized by \textit{ip}, except for those that descend from any element in \textit{Exc}. 
Finally, a privacy preference allows data owners to specify what  cannot be derived from their data by the analytics processes. This is specified through a \textit{category derivation constraint -- cdc}, that specifies a set of data categories, denoted as  \textit{Cdc}, that cannot be derived using the data.\footnote{In what follows, we will use the dot notation to refer to specific components of $jac$ and $ip$ elements.}


A privacy preference is therefore formally defined as follows:

\begin{definition} {\bf (Privacy preference)}\label{def:pp} 
A privacy preference $pp$ is a tuple $\langle${$\alpha$, \textit{consumer}, \textit{ip}, \textit{jac}, \textit{cdc}}$\rangle$, where $\alpha$ is an attribute of a data stream generated by a smart object, \textit{consumer}  specifies the set of data consumers authorized to access $\alpha$,  
\textit{ip} specifies the intended purposes for which $\alpha$ can be collected and used, \textit{jac} specifies a joint access constraint, whereas \textit{cdc} is category derivation constraint specifying a set of data categories  which cannot be derived from $\alpha$.
\end{definition}

A summary of the notations related to privacy preferences is given in Table \ref{tab:ppElements}.

\begin{table}[H]
    \centering
    \scriptsize 
    \begin{tabular}{|c|c|}
        \hline
        {\bf  Element} & {\bf Abbreviation} \\
		\hline
        \textit{Consumer ids} & \textit{consumer} \\
        \hline
        \textit{Intended Purpose} & \textit{ip} \\
        \hline
        \textit{Allowed intended purposes of ip} & \textit{ip.Aip} \\
        \hline
        \textit{Prohibited intended purposes of ip} & \textit{ip.Exc} \\
        \hline
        \textit{Joint access constraint} & \textit{jac}  \\
        \hline
        \textit{Allowed data categories of jac} & \textit{jac.Adc} \\
        \hline
        \textit{Prohibited data categories of jac} & \textit{jac.Exc} \\
        \hline
        \textit{Intended purposes for jac} & \textit{jac.ip} \\
        \hline
        \textit{Category derivation constraint} & \textit{cdc} \\
        \hline
    \end{tabular}
\caption{Privacy preference's notation} \label{tab:ppElements}
\end{table}


\begin{example}\label{exp:PPCharlotte}
Let us consider the smart home scenario given in Example  \ref{exp:smartObjects} and let us
 suppose that a user, say Charlotte, has installed this smart home system in her apartment and that she wants to specify  a privacy preference \textit{pp} for the  temperature data sensed by the system. In particular, 
Charlotte wants to share her data only with \textit{Smart Home Company} and she wants to release the sensed data only for the administration of the heating system. Moreover, she allows temperature data to be joined with  generic information, as well as health data of her smart watch (e.g.,  body temperature) as this  could  help her to properly customize  room temperature. 
Thus, she agrees that  temperature data is joined with generic information and health data for administration purpose. However,  she prefers that from these data it is not possible to derive sensitive information, such as her current location in the apartment or the number of persons in the room. To model these requirements, Charlotte can specify  \textit{pp}  as:  \textit{$\langle$\textit{temperature}, \{smart-home-company\}, $\langle$\{admin\}, $\emptyset\rangle$, $\langle$\{generic, health\},$\emptyset$, $\langle$\{admin\},$\emptyset\rangle\rangle$, \{sensitive\}$\rangle$}.
\end{example}

As a final note,  it is important to highlight  that the additional elements that we have introduced to cope with the new issues that IoT scenario poses, (i.e., consumer identities, joint access constraints, and category derivation constraints), are not mandatory. This provides flexibility to vendors to decide   whether to  adhere only to  common privacy practices  (e.g., supporting privacy preferences with purpose, retention, and third party elements) or  to exploit  the extended privacy preferences so as to provide users with a more complete control on their personal data.

\subsection{Privacy Preference Derivation}
An interesting feature of the model in \cite{ourPaper} is its ability to associate privacy preferences with new data generated by smart objects. These new privacy preferences are defined  by combining the privacy preferences specified by each owner of smart objects sensing the data involved in the derivation process.   
Privacy preferences are combined by taking the most conservative approach so that the resulting privacy preference satisfies the constraints specified in all the  privacy preferences associated with smart objects involved in the data fusion process. 
More precisely, the composed privacy preference is defined by taking the intersection of the \textit{consumer, jac.Adc} and \textit{ip.Aip} elements, and the union of the \textit{jac.Exc, ip.Exc} and \textit{cdc} fields of the privacy preferences involved in data fusion. 
Formally a  composed privacy preference is defined as follows:
\begin{definition}{\bf (Composed privacy preference)} \label{def:composedpp}
Let $\alpha$ be an attribute belonging to the output stream generated by a smart object, and let $At$ be the set of attributes from which $\alpha$ is derived. The privacy preference $pp$ = $\langle${$\alpha$, \textit{consumer}, \textit{ip}, \textit{jac}, \textit{cdc}}$\rangle$ for $\alpha$ is derived as follows:

\begin{itemize}
\item \textit{consumer} is defined as the intersection of the ids in the \textit{consumer} field with all privacy preferences associated to all attributes in $At$, that is:  $consumer=\bigcap_{a\in At}a.pp.consumer$

\item  The intended purpose component $ip$ is defined as:

-- $Aip$ (allowed intended purposes) is the intersection of the purposes implied by the \textit{ip}'s allowed intended purposes of all attributes in \textit{At}, that is: 
$Aip=\bigcap_{a\in At}$ $\bigcup_{p \in a.pp.ip.Aip} p^\downarrow$, where $p^\downarrow$ denotes a set composed of \textit{p} and all purposes descending from \textit{p} in the purpose tree. 

--  $Exc$ (prohibited purposes) is the union of the \textit{ip}'s prohibited purposes of the privacy preferences specified for \textit{At}'s attributes, that is:  $Exc=\bigcup_{a\in At} a.pp.ip.Exc$

\item  The joint access constraint \textit{jac}  is defined as:
--  $Adc$ (allowed data categories) is the intersection of the \textit{jac}'s allowed data categories of all attributes in $At$, that is:  $Adc=\bigcap_{a\in At}\bigcup_{c\in a.pp.jac.Adc} c^\downarrow$, where $c^\downarrow$ denotes the set composed of c and all categories that descend from \textit{c} in the data category tree.

--  $Exc$ (exception) is the union of the \textit{jac}'s exceptions specified for all attributes in $At$, that is: \\ $Exc$=$\bigcup_{a\in At}a.pp.jac.Exc$;

-- $ip$ (\textit{jac} intended purposes) is defined following the same criteria of the preference intended purposes, that is:  $ip.Aip$=$\bigcap_{a\in At}\bigcup_{p \in a.pp.jac.ip.Aip}$ $p^\downarrow$, and  $ip.Exc=\bigcup_{a\in At} a.pp.jac.ip.Exc$.
\item The prohibited data categories \textit{cdc} are defined as the union of prohibited data categories specified in the privacy preferences of all attributes in $At$, that is:  $cdc=\bigcup_{a\in At}a.pp.cdc$

\end{itemize}
\end{definition}


\begin{example} \label{exp:composedPP}
Let  us consider the privacy preference described in Example \ref{exp:PPCharlotte} and let us further suppose that Charlotte also wants to specify  a privacy preference for the electricity usage data. In particular,   Charlotte wants to share her data with the \textit{Electricity Company} and the \textit{Smart Home Company}. Moreover, she wishes to: i) release the sensed data only for the administration of the electricity usage system, and ii) to join them with generic information for administration purpose, whereas iii) she does not pose any condition on what can't be derived from her data. Thus,  she also specifies the following privacy preference: \textit{$\langle$\textit{electricityUsage}, \{smart-home-company, electricity-company\}, $\langle$\{admin\},$\emptyset\rangle$, $\langle$\{generic\},$\emptyset$, $\langle$\{admin\}, $\emptyset\rangle\rangle$,  $\emptyset\rangle$}.

Let us  assume that the smart home network executes an equi-join on streams containing  the electricity usage and temperature data generated by  Charlotte's smart home system on $id_{room}$ attributes, which we assume to have in both the two streams. Moreover, we assume that $id_{room}$  has associated the same privacy  preferences of temperature and electricity usage in both streams. 
The composed privacy preference for $id_{room}$ in the stream containing  temperature data is: 

\textit{$\langle$\textit{$id_{room}$}, \{smart-home-company\}, $\langle$\{admin\}, $\emptyset\rangle$, $\langle$\{generic\}, $\emptyset$, $\langle$\{admin\}, $\emptyset\rangle\rangle$, \{sensitive\}$\rangle$}.

This new  privacy preference states that  information on $id_{room}$ can be released only for administration purpose to the Smart Home Company, can be joined with generic data for administration purpose, whereas  sensitive information cannot be derived from this data.
\end{example}

\section{Decentralized privacy preference enforcement} \label{sec:enforcement}
In order to assure that  user privacy preferences are taken into consideration during  data usage and processing, there is the need of an enforcement mechanism whose primary goal is to verify  whether the privacy practices adopted by the entity wishing to consume the data, aka the consumer, satisfies  data owners'  privacy preferences.
To make this compliance check automatic, we assume that consumer  privacy practices are encoded into privacy policies, that is, a set of statements on how  the consumer will use the users' data. Typically, a privacy policy specifies:  the consumer identity (id), data attributes that consumer wishes to consume, purpose of the data usage, release of the data to third-parties and retention of the data in the consumer information system.\footnote{Privacy policies can be specified, as an example, by using XACML (https://docs.oasis-open.org/xacml/3.0/privacy/v1.0/xacml-3.0-privacy-v1.0.html)}
In what follows, we focus on the purpose and consumer id components of a privacy policy/preference and the related checks, in that they are the most important information used by the enforcement monitor. 
The enforcement mechanism has to verify whether the declared purpose of data usage and id of the consumer in the consumer's privacy policy is compliant with the  privacy preferences specified  by the users owning the smart object devices generating the data.

We handle compliance check through an aposteriori approach, that is, we perform compliance check only when data are going to be shared with consumers.
The main motivation is that leaving compliance check to  consumer SOs increases the possibility for data to be processed and thus to be eventually consumed according to user privacy preferences.
For instance, let us consider the equi-join operation performed in Example \ref{exp:composedPP}. 
Let us assume that Charlotte's privacy preferences for electricity usage data  only allow joint access with position data (e.g., coordinates of the individual), and let us also assume that joined streams have further attributes, in addition to electricity usage and temperature data.
In such case, the equi-join operation in Example \ref{exp:composedPP} is not compliant with Charlotte's privacy preferences.
Performing a compliance check at this point would prevent to generate the joined stream, and further data flows would be cut. In contrast, by an aposteriori compliance check approach attributes in the joined streams can be further processed in the smart object network and consumers can consume the data that would not violate the privacy preferences of the users.


\begin{figure} [h]
	\centering
	\includegraphics[width=85mm]{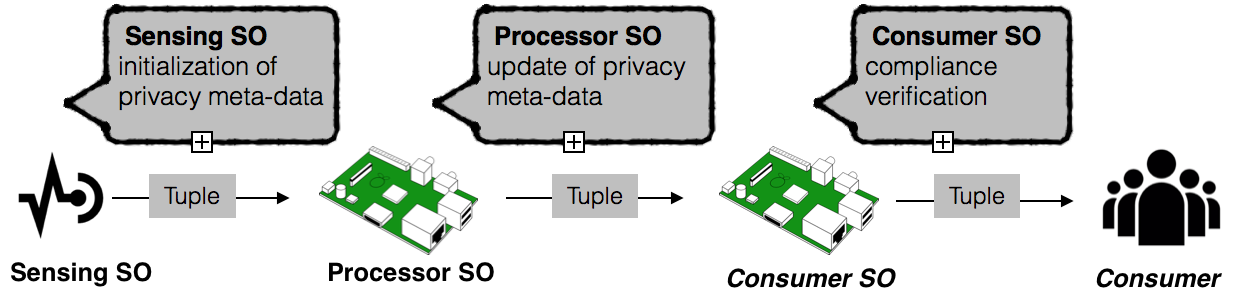}
	\caption{Privacy enforcement by SO roles \label{fig:SOenforcement}}
\end{figure}

In order to push   privacy preference enforcement at object level, we have to first complement  the processed data with {\em privacy meta-data} that specify, for each piece of sensed or newly created data,  all information needed to enforce  privacy preferences (see Figure \ref{fig:SOenforcement}). As an example,  the sensing smart objects have to be able to   associate with each  piece of sensed data the corresponding  privacy preferences specified by object owners. Moreover, processor SOs have to be able to locally generate additional privacy preferences for the newly created or modified data. This process requires to make  processor SOs aware of how  the new data have been created (e.g., which are the operations, the input data involved in the operations, as well as, the corresponding privacy preferences). Finally, smart objects playing the role of consumer SOs have to be able to locally perform   compliance check.

\subsection{Privacy meta-data} \label{sec:peas}
 To model  additional  privacy information, we need to add, for each attribute contained into the  original schema, three additional attributes, namely {\em category}, {\em pp} (privacy preferences), and {\em history}. 
 
Let us start discussing  the  {\em category} attribute. 
The category associated with an attribute is not a static information, as  data associated with an attribute  might be fused with other information, or simply  modified, with the result of a change of its content and thus of its associated data categories.  To cope with these dynamic aspects,  we need to trace, for each attribute,  its updated set of data categories, based on its current content.
Secondly,  in order to make a consumer SO able to perform  privacy preference enforcement, it has to know the privacy preferences associated with each attribute. Such information is  encoded in the additional  {\em pp} attribute.  
Moreover, information about every operation performed on data has to be documented alongside the data itself.  Indeed, by design,  compliance check is only performed  by consumer SOs, just before data are sent to the consumer (see Figure \ref{fig:SOenforcement}).  Processor SOs are only in charge of executing data processing and privacy meta-data generation, without caring about compliance checks. As such,  it might be possible that  in a smart object network, a processor SO performs an operation, e.g., a join, that breaks a $jac$ or $cdc$ condition. In order to make a consumer SO able to verify  if operations performed by all processor SOs satisfy  $jac$ and $cdc$ constraints, it has to know all operations performed on each piece of data (aka each attribute). At this purpose, we introduce, for each  attribute in the stream schema, a new attribute, denoted as {\em history}, which contains a list of history entries, one for each operation performed on the corresponding attribute.

This privacy meta-data are  directly encoded into the original data stream schema, obtaining thus a new schema,  called {\em Privacy-Enhanced Attribute Schema} (PEAS), formally defined as follows:

\begin{definition}\textbf{(Privacy-Enhanced Attribute Schema).}\label{peas}
Let $S$=($A_1,\dots,A_n$) be the schema of a data stream, where  $A_1,\dots$,$A_n$ are the data stream attributes.
The Privacy-Enhanced Attribute Schema of $S$ is defined as $\overline{S}$=($\overline{A}_1,\dots$,$\overline{A}_n$), such that $\overline{A}_j$= ($A_j$, $pp$, $category$, $history$), $\forall j\in{1,\dots,n}$, where:
\begin{itemize}
    \item $pp$ is the privacy preference associated with attribute $A_j$, as defined in Definition \ref{def:pp};
    \item $category$ is the set of data categories associated with the current value of attribute $A_j$;
    \item $history$ is a list \{$HE_1$,$\dots$,$HE_m$\}, containing an element for each operator Op, $A_j$ has undergone so far. $HE_i$=($AC_i$, $RC_i$), for $\forall i\in{1,\dots,m}$, where: \\
      -- $AC_i$  is the set of categories associated with attributes on which Op is executed  for deriving  $\overline{A}_j$; \\
      -- $RC_j$ is the set of data categories associated with  $\overline{A}_j$ after the execution of Op; 
\end{itemize}

\end{definition}

As depicted in Figure \ref{fig:SOenforcement}, privacy meta-data are initialized by  sensing SOs, which collect the raw  data and generate the corresponding PEAS schemas. In particular, at this stage in the PEAS schema  the $history$ field is initialized as empty, whereas the $pp$ and $category$  fields contain privacy preferences and associated data category information defined by smart object owners. Processor SOs also modify attributes of PEAS schema. In particular, as it will be discussed in Section \ref{PESderivation}, for each processed attribute,  they have to  update the  $category$ field, based on current value of  the corresponding attribute as well as the $history$ field.

\begin{example}\label{exp:PEASCharlotte}
Let us  refer again to Example \ref{exp:PPCharlotte}, and, in particular, to   attribute $att$ containing the temperature data of Charlotte's smart home. Recalling the privacy preferences  in Example \ref{exp:PPCharlotte}, and supposing that the current temperature is 24.4 Celsius degree,
the initial PEAS generated by the sensing SO, is the following:\\
-- $att$ = \{24.4\},\\
-- $att.pp$ = $\langle$temperature, \{smart-home-company\}, $\langle$\{admin\}, $\emptyset\rangle$, $\langle$\{generic, health\}, $\emptyset$, $\langle\{admin\},  \emptyset\rangle\rangle$, \{sensitive\}$\rangle$, \\
-- $att.category$ = \{generic\}, \\
-- $att.history$ = \{$\emptyset$\}. \\  

\end{example}

\subsection{PEAS generation}\label{PESderivation}
For each sensed data,   the sensing SO has to initialize the  proper PEAS, with the privacy preferences specified by sensing SO's owner, the initial data category of the sensed data, and an  empty {\em history} field. 
Additionally,  after each operation, processor SOs should derive  the privacy preferences of the eventually newly created attribute content, as well as  update the \textit{category} and \textit{history} fields of each attribute involved in the operation. Hence,  we assume that processor SOs are complemented  with additional logic, implementing  PEAS update.

Since  PEAS generation in sensing SOs is quite straightforward, in that it mainly consists in initializing $pp$ and $category$ attributes with predefined values, in the following we focus on PEAS update algorithms. PEAS update  is mainly driven by the operations performed by  processor SOs. As such,  we define algorithms  for all query operators except for the selection operator, as, by definition, this  operator does not generate any  new data, as such, no modification of the PEAS is needed.  The algorithm receives as input the performed SQL operator $Op$ (e.g., avg, sum, $\times$, etc.) and input streams, used by the corresponding operation. In particular, we model  an SQL operator based on operations it performs on the attributes. For example, the  join of  streams $S1(a,b,c)$ and $S2(a,d,e)$  on attribute $a$ is modelled as two parameters: \{join, (S1.a, S2.a)\}, that is, the involved attributes and the performed operation.  Since, an SQL operator  $Op$ might perform more than one operation (see, for instance, the $\Pi$ operator in the following Example  \ref{exp:projection}), we formally define $Op$ as a set \{Pa$_1$,$\dots$,Pa$_l$\},  where every  Pa$_i$ models a single operation. As  such, Pa$_i$= \{\textit{Attributes,$fn$\}}, $\forall$ $i\in\{1,\dots,l\}$, where  $Attributes$ is a set of attributes given as input to function $fn$  when  operation $Pa_{i}$  is executed. 

\begin{example}\label{exp:projection}
Suppose that the stream resulting from the  equi-join operation illustrated in Example \ref{exp:composedPP} also contains \textit{humidity} data, thus resulting in \textit{S(temperature, electricityUsage, humidity)}. 
Let us assume to perform on this joined stream a projection $\Pi{_1}$ computing two values: the first  obtained as multiplication of  \textit{temperature} and \textit{electricityUsage} to estimate the number of people in the smart home,\footnote{For simplicity, we assume that this is possible by multiplying electricity and temperature usage.}  and the second as execution of $f()$ that takes as input $temperature$ and $humidity$ and returns 
the air-quality information. $\Pi{_1}$ is thus modeled as: ${\Pi{_1} = \{Pa_1,Pa_2\}}$, where Pa$_1$= \{\{temperature, electricityUsage\}, $\times$\} and Pa$_2$= \{\{temperature, humidity\}, $f()$\}. 
\end{example}

PEAS generation algorithm exploits  function \textit{derivePP()}, that  derives new composed privacy preferences, and function \textit{createHistoryEntry()}, which creates a new \textit{history} entry, based on the parameters of the performed operation. Finally, function \textit{deriveDC()}  determines the category of the new generated data given in  input. To implement  such a function, we assume the presence of a set of inference rules, called {\em derivation paths}. Each derivation path associates with  a query operator and the set of categories of the data on which the operator is executed, the category of the  data resulting from the operator execution. We therefore assume that derivation paths are present in every processor SO, so derived data category can be locally computed.

\begin{example}\label{exp:dp}
Let us give an example of derivation path, by referring to the smart home scenario illustrated in Figure \ref{fig:smarthome} and by assuming that also humidity information can be sensed. 
The derivation path $dp_1$= $\langle\{$temperature$, $electricityUsage$\}$, $\times$, $\Pi$ , sensitive $\rangle$,  specifies that the knowledge of temperature and electricity usage allows, through the  $\Pi$  and $\times$ operators, the derivation of the number of people  inside the smart home.
In contrast, the derivation path $dp_2$= $\langle\{$temperature$, $humidity$\}$, $f()$, $\Pi$, air-quality $\rangle$
specifies that the knowledge of temperature and humidity usage allows derivation of air quality information, through operator $\Pi$ and function $f()$.
\end{example}

For PEAS derivation, Algorithm \ref{alg:unique} takes as input tuples $s_1$ and $s_2$ (if operation is not $\Join$ then $s_2$ will be empty), and parameters modeling operation $\Pi$, $\Sigma$ or $\Join$. The algorithm returns an updated tuple, containing new attribute(s) generated by execution of the corresponding operation given in input and containing the updated $pp$, $category$ and $history$ fields associated with each attribute involved in the operation. In performing PEAS update for $\Join$ operation, attributes of tuples $s_1$, $s_2$ not used by $\Join$ remain unchanged, and by the nature of $\Sigma$  (such as avg(), sum(), etc.), only one parameter will be present in $Op.Parameters$. The algorithm exploits previously discussed functions: \textit{derivePP()} and \textit{createHistoryEntry()}. 


\removelatexerror
\noindent\begin{minipage}{85mm}
\renewcommand\footnoterule{}  
\begin{algorithm} [H]
\caption{{\em PEAS Derivation}(${Op},{s_1}, {s_2}$) }\label{alg:unique}\footnotetext{For $s_{new}$ and $att_{new}$, dot notation is used in accessing their elements specified by Definition \ref{peas}.}
\scriptsize{
Let $s_{new}$ be a new stream, initialized with an empty schema\;
Let $refAttSet$ be the set of attributes, initialized to be empty\; 
\uIf{Op = {$\Sigma$ or $\pi$}}
{
	\For{ each $Pa\in Op.Parameters$}
  {
      Let $att_{new}$ be an attribute, initialized to be empty\;
      Let $newData$ be the result of  operation $Pa$\;
      $refAttSet$ = Pa.Attributes\;
      $att_{new}$.name = $Pa.name$\;
      $att_{new}$.data = newData\;
      $att_{new}$.category = deriveDC(Op, $refAttSet$)\;
      $att_{new}$.PP = derivePP(Op, $refAttSet$)\; 
      $att_{new}$.History =  $\bigcup_{att\in refAttSet} att.History \bigcup$ createHistoryEntry($\bigcup_{a\in refAttSet} a.Category$, $att$.category)\;
      $s_{new}$.Attributes = $s_{new}$.Attributes $\cup$ $att_{new}$\; 
  }
}
\uElseIf{Op = $\Join$}{
  Let Op.Parameters=\{$a_{1}$, $a_{2}$\} be the two attributes used in the Join statement\;
  Let $s_{1}.Attributes$ and $s_{2}.Attributes$ be the set of attributes contained in $s_{1}$ and $s_{2}$ schema, respectively\;
  Let $attributeSet$ be a set of attributes, initialized to be empty\;
  
  $refAttSet$ $=a_{1}$ $\cup$ $a_{2}$\;
  $attributeSet$ = $s_{1}.Attributes$ $\cup$ $s_{2}.Attributes$ $\setminus$ $refAttSet$\;
  $a_{1,2}$.PP = derivePP(Op, $refAttSet$)\

  $a_{1,2}$.History = $a_{2}$.History $\bigcup$ createHistoryEntry($refAttSet$.category, $a_{1}$.category | $a_{2}$.category )\; 

    
  $s_{new}$.Attributes = $attributeSet$ $\bigcup$ $refAttSet$\;
}

{\bf Return} $\langle s_{new}\rangle$\;
}

\end{algorithm}
\end{minipage}

\begin{example}\label{exp:join}
Let us give an example of PEAS derivation by Algorithm \ref{alg:unique} for the join of temperature and electricity usage data described in Example \ref{exp:composedPP}. Let us assume that the PEAS of electricity usage data has been created with the same logic given in Example \ref{exp:PEASCharlotte}.
First, Algorithm \ref{alg:unique} stores the parameters used in the join operation into \textit{refAttSet}. 
All attributes of the two streams except parameters used in the join operation are stored as \textit{attributeSet} to be included in the resulting stream without any modification. 
Then, the algorithm performs composition of privacy preferences for attributes in \textit{refAttSet} and updates attributes in \textit{refAttSet} with newly derived privacy preferences.
Then, the  new history entry \textit{newHE} is created by function \textit{createHistoryEntry()}, and added to the history fields of both  the attributes used as parameters in the operation.
Finally, the union of the attributes used as parameters of the operation and \textit{attributeSet} are added to the resulting stream. The updated PEAS for the  temperature attribute is therefore: \\
-- $att$ = \{24.4\},\\
-- $att.pp$ = $\langle$temperature, \{smart-home-company\}, $\langle$\{admin\}, $\emptyset\rangle$, $\langle$\{generic\}, $\emptyset$, $\langle\{admin\},  \emptyset\rangle\rangle$, \{sensitive\}$\rangle$, \\
-- $att.category$ = \{generic\},  \\  
-- $att.history$ = \{\{generic, generic\}, \{generic\}\}. 

\end{example}

\subsection{Compliance verification}\label{complianceAnalysis}
Compliance verification is the operation of verifying if constraints specified by  data privacy preferences are satisfied by the privacy policy of the consumer. These checks are performed by consumer SOs on every piece of data  passing through them. More formally, a consumer SO receives as input a stream $S$ and returns as output a stream $\overline{S}$, containing only those  tuples whose attributes' values satisfy privacy preferences of all the owners involved in the generation of attributes' contents. Compliance check implies three main steps: 
$(i)$ checking compliance of the consumer's identity (i.e., $consumerID$) with the allowed consumers (i.e., $consumer$) specified in the data privacy preference;
$(ii)$ checking compliance of the purpose (i.e., $ap$) of the consumer with the intended purposes (i.e., $ip$) specified in the data privacy preference;
and $(iii)$ checking compliance of the operations performed on the data (e.g., on each single attribute of the input stream)  with constraints imposed by the joint access constraint $jac$ and the category derivation constraint $cdc$.
 
Step $(i)$ is straightforward in that it should only be checked that    $consumerID$ specifies the identity of a consumer allowed to consume attribute $A$, that is,  $consumerID \in A.pp.consumer$.
Regarding  step $(ii)$, this check is satisfied if $ap$ specifies an allowed  purpose, that is, a purpose contained in $ip$ or in the set of allowed purposes implied by $ip$, if purposes are organized into a tree. More formally, denoting  with  $\vec{ip}$  the set of purposes implied by \textit{ip},  $ap$ complies with $ip$ iff  $ap\in A.pp.\vec{ip}$.  

For step $(iii)$, let us first discuss compliance of joint access constraint $jac$ for  a given attribute $A$ with privacy preference $pp$. $Jac$ states which portions of user's personal data can be  combined/aggregated with other data, and  with which consumer purpose, as described in Section \ref{sec:privacyModel}. 
The set of categories implied by $jac$ denoted as $\vec{jac}$, is composed of all categories that descend from the categories in $jac.Adc$ which do not descend from any category of $jac.Exc$.
We say that the $jac$ of attribute $A$ is satisfied iff: $(i)$ all the accessed data categories in every history entry of $A$ belong to the set of categories implied by $\vec{jac}$, and $(ii)$ $ap$ complies with the intended purpose component of $jac$. 
Formally:

\begin{definition} \textbf{Jac compliance.}\label{jacCompNew}
Let $pp$ be a privacy preference specified for attribute $A$. Compliance of the $pp.jac$ component is satisfied iff: $ap \in jac.\vec{ip}$ $\wedge$ $\forall ${HE}$\in A.history$, $\forall c \in ${HE}$.\vec{AC}$  $\rightarrow$ $c \in A.pp.\vec{jac}$.
\end{definition}

Let us now discuss compliance of category derivation constraint $cdc$ for a given attribute $A$ with privacy preference $pp$, which specifies the data categories that cannot be derived from the processing of attribute $A$. 
The set of data categories which on the basis of $cdc$ cannot be derived denoted as $\vec{cdc}$, is composed of all categories that specialize those referred to within $cdc$.
Similarly to what happened for the $jac$ component, history of the resulting attribute must be checked, since derivation of new attributes  may be performed several times, and every derivation process has to comply with privacy preferences of the attribute. Therefore, we say that $cdc$ of  attribute $A$ is satisfied iff all the data categories in every history entry of $A$ does not belong to the set of categories implied by $\vec{cdc}$. 
Formally:

\begin{definition} \textbf{Cdc compliance.}\label{cdcCompNew}
Let $pp$ be a privacy preferences specified for attribute $A$. Compliance of  $pp.cdc$ is satisfied iff: $\forall ${HE}$\in A.history$,  $\nexists c \in ${HE}$.\vec{RC}$  $\wedge$ $c \in A.pp.\vec{cdc}$.
\end{definition}

Consumer SOs perform compliance check for any attribute in every tuple before  sharing it with consumer. Let us present and describe our algorithm for checking compliance. As previously discussed, $jac$ and $cdc$ compliance have to be ensured for every entry in the $history$ field of PEAS. Algorithm \ref{alg:CC} takes as input a tuple $s$, the id of the consumer $consumerID$, and the access purpose of the consumer $ap$.  The algorithm returns an updated tuple, containing attribute(s) that complies with the privacy preferences of the involved users and access purpose of the consumer.

\noindent\begin{minipage}{85mm}
\renewcommand\footnoterule{}  
\begin{algorithm} [H]
\caption{{\em complianceCheck}(${s}, {consumerID}, {ap}$)}\label{alg:CC}\footnotetext{For $s_{new}$ and $att_{new}$, dot notation is used in accessing their elements specified by Definition \ref{peas}.} 
\scriptsize{

Let $s_{new}$ be a stream, initialized as empty\;
\For{ each $A\in S$}
{
    Let $cdcFlag$ and $jacFlag$  be boolean variables, initialized as true\;
    Let $ipFlag$ and $jacIpFlag$  be boolean variables, initialized as false\;
    
  \uIf{$(consumerID\in A.pp.consumer)$}
  {   
  
    \If{$(ap\in A.pp.\vec{ip})$}
        {$ipFlag$ = true\;}
     
    \If{$(ap\in A.pp.jac.\vec{ip})$}
        {$jacIpFlag$ = true\;}

    \For{ each $HE\in A.History$}
    {
        Let  $accessedCategories$ = ${HE}$.AC\;  
        Let  $resultCategories$ = ${HE}$.RC\; 
        Let  $jacDataFlag$ be a boolean variable, initialized as true\;
        
        \For{ each $cat \in accessedCategories$}
         {
            Let $flag$ be a boolean variable, initialized as false\;
            \If{$cat \in A.pp.\vec{jac}$}
                {$flag$ = true\;}
            $jacDataFlag$ = ${jacDataFlag\wedge flag}$\;
        }  
                
        \For{ each $cat2 \in resultCategories$}
         {
            Let $flag2$ be a boolean variable, initialized as true\;
            \If{  $cat2 \subseteq A.pp.\vec{cdc}$  }
                {$flag2$ = false\;} 
            $cdcFlag$=$cdcFlag\wedge{flag2}$\;  
         }           
    }

    $jacFlag$ = jacDataFlag$\wedge$jacIpFlag\;
    	
    \If{$jacFlag$=true $\bigwedge$ $cdcFlag$=true $\bigwedge$ $ipFlag$=true}
    {
        $S_{new}$ = $S_{new}$ $\bigcup$ $A$\;
    }
  }                
}
{\bf Return} $\langle$ $S_{new}$ $\rangle$\;
}
\end{algorithm}
\end{minipage}


\begin{example}
Let us suppose that a processor SO performs the equi-join and projection $\Pi_{1}$ given in Example \ref{exp:join}. Moreover, suppose to have the derivation paths  presented  in Example \ref{exp:dp}. \\
Algorithm \ref{alg:CC} first checks compliance of the consumer (line 5). Let us assume that the consumer entity is \textit{smart-home-company}. In this case, the check returns true.
Then, Algorithm \ref{alg:CC} checks compliance of  \textit{peopleCount.pp.ip} and \textit{peopleCount$.$pp.jac.ip}  with the given consumer access purpose $ap$ (lines 6 and 8). Let us assume that the consumer has specified \textit{admin} as access purpose.  In this case, the check  returns true as both the purposes are \textit{admin}, hence \textit{ipFlag} and \textit{jacIpFlag} are set to true. 
Then, for both $jac$ and $cdc$, Algorithm \ref{alg:CC} checks compliance of every history entry $HE$. Since $peopleCount.history$ has two elements, the check is performed twice (line 10).  In both of them, the algorithm checks that every category $cat$ in  $HE.AC$  belongs to $peopleCount.pp.\vec{jac}$ (line 16). In our example, as \textit{peopleCount.pp.jac} allows joint access with data belonging to category \textit{generic}, $jacDataFlag$ will remain true for both of the history entries of $peopleCount.History$. In the next step, the logical conjunction of $jacDataFlag$ and \textit{jacIpFlag} is done, to get $jacFlag$. Since  both of them are true, $jacFlag$ will be true. Similarly, the algorithm checks that every category $cat2$ in  $HE.RC$ does not belong to $peopleCount$$.pp.\vec{cdc}$ (line 21). For the first history entry, $cdcFlag$ will remain true, as category in $HE.RC$ is not sensitive. For the second history entry, $cdcFlag$ will change to false, since the category in $HE.RC$ is sensitive. Finally, since $cdcFlag$ is false (line 25), attribute $peopleCount$ is not  added to the final schema, hence it will not be shared with the consumer.
\end{example}

\section{Experiments}\label{sec:experiments}

In order to evaluate the overhead of the proposed mechanism, we have implemented several scenarios introduced in what follows.

\subsection{Experimental scenarios} \label{sec:expScenarios}
We developed four experimental scenarios to estimate the  overhead at smart object  and  at network level.

\subsubsection{Processor Smart Objects \rom{1} }\label{expSce1} 
In this scenario, we consider  processor SOs  of second   smartness level, able only to perform data selection and projection. In particular, we assume that processor SOs \rom{1} are physically manufactured low capability devices, with hard-coded logic.To simulate this low capability devices, we have implemented  smart objects  using Freescale FRDM-K64F\footnote{developer.mbed.org/platforms/FRDM-K64F/} with \textit{mbed} operating system.\footnote{www.mbed.com/en/platform/mbed-os/} This device has   256 KB RAM and ARM Cortex-M4 core.  Coding has been done using C++. Due to limited capabilities of the FRDM platform, we were not able to use real world data. As such, in this scenario, we used  synthetic data, randomly  created on the FRDM platform. We generate a tuple per second, containing a unique attribute of float data type, with randomly assigned value.

\subsubsection{Processor Smart Objects \rom{2} }\label{expSce2} 
In this scenario, we assume that the processor SO   is a device with third smartness level, able  to perform, in addition to selection and projection operations, also window-based operations, such as join and aggregation. This processor SO \rom{2} has been implemented on  a Raspberry Pi 3 Model B with \textit{Raspbian} operating system, with 1GB RAM and a 1.2 GHz 64-bit quad-core ARMv8 CPU. The implementation has been done with Java SE Development Kit 8u73. Moreover, to implement  into processor SOs \rom{2} an SQL engine able to manage window-based operations, we exploit Esper.\footnote{www.espertech.com/esper} We make use of the IoT-Lab testbed\footnote{www.iot-lab.info} data, which  provides a very large scale infrastructure suitable for testing small wireless sensor devices and heterogeneous communicating objects. Each node of IoT-Lab testbed produces tuples containing a unique attribute  containing data sensed by that node.  
In particular, IoT-Lab provides three hardware platforms: WSN430, M3 Nodes, and A8 Nodes.  Our experiments have been run with data transferred from M3 Nodes, where lumen level data, represented as float values, are sensed from light-to-digital sensor ISL29020.\footnote{www.digikey.com/catalog/en/partgroup/isl29020/14151}

\subsubsection{Consumer Smart Objects}\label{expSce3} 
We implemented a smart object with third smartness level, able  to check privacy preference compliance. We adopt the same simulation and data generation strategies used for processor SOs \rom{2} (see Section \ref{expSce2}), where compliance verification algorithm has been implemented as an additional Java function.

\subsubsection{Smart Objects network}\label{expSce4}
\begin{figure} [H]
	\centering
	\includegraphics[width=0.5\textwidth]{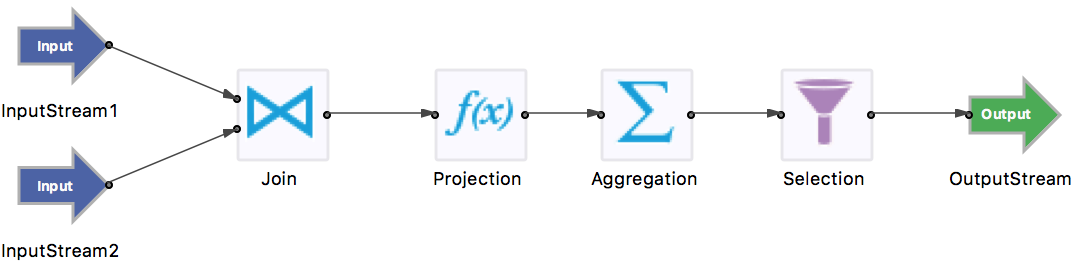}
	\caption{Graph-based SQL query \label{fig:streambase}}
\end{figure}

We recall that the aim of this experiment  is to measure the overall time overhead and bandwidth utilization  implied by our enforcement mechanism on a network of smart objects. To simulate this scenario, we should  implement a real network of smart objects, where each object is  modified so as to embed the PEAS update algorithm as well as the compliance checks. 
However, the lack of standard platforms  on which injecting our modified  smart objects makes hard to deploy this setting.  To overcome this issue, we decided to simulate a network of smart objects by  leveraging on  Complex  Event  Processing  (CEP) systems \cite{CEPexp}. We exploit the graph-based SQL modelling of CEP to simulate a network of objects, where every single operation in the CEP query is interpreted as  a distinct smart object.  As an example, the graph-based SQL query in Figure \ref{fig:streambase} can be interpreted as a network of two sensing SOs (i.e., the two INPUT operators), four processor SOs (i.e., the join, projection, aggregation  and selection operators), and one final  OUTPUT  operator. 

We adopted  Streambase as CEP platform, as this product allows us to associate with  an SQL operator an additional function, running a Java code,  that  will be executed by Streambase every time the operator  processes an incoming tuple.   As such, with Java SE Development Kit 8u73, we  defined a set of functions implementing  the   PEAS update  algorithms (cfr. Section \ref{sec:enforcement}). These functions have been properly associated  with    INPUT, Join, Selection, Projection, and Aggregation operators of Streambase. Following this design, the  compliance verification algorithm should have been  associated on the  OUTPUT operator. However, since Streambase does not allow this,   we implemented a new Streambase operator, to which the function implementing the compliance verification algorithm has been  associated. This new defined operator is inserted just before the  OUTPUT operator finalizing the query. We have exploited the feed simulation tool from Streambase Studio, that automatically generates and passes test data at specified rates to INPUT operators. Data rate has been fixed to 100 tuples per second for each stream. Number of attributes in each tuple has been regulated according to experiment settings (see discussion on query complexity  in Section \ref{exp2}). 

\begin{table}[t]
    \centering
    \scriptsize 
    \begin{tabular}{|c|c|c|c|c|}
        \hline

        {\bf Tested}  & {\bf Proc.} & {\bf Proc. } & {\bf Consumer } & {\bf Smart} \\
      
        {\bf dimensions}  & {\bf SO \rom{1}} & {\bf SO \rom{2}} & {\bf  SO} & {\bf Object} \\
        {\bf }  & {\bf  } & {\bf } & {\bf  } & {\bf network} \\
        \hline
        \textit{Varying } & &  &  & \\
        \textit{complexity } & \textit{ $\checkmark$ } & \textit{$\checkmark$ } & \textit{ $\checkmark$ } & \textit{ $\checkmark$ } \\
        \textit{of PP} &  &  &  &  \\
        \hline
        \textit{Varying}   &  &  &  &  \\
        \textit{query }  & \textit{ $\checkmark$ } & \textit{ $\checkmark$ } & \textit{ $\checkmark$ } & \textit{ $\checkmark$ *}\\
        \textit{complexity}   &  &  &  &\\\hline
        \textit{Varying }  &  &  &  & \\
        \textit{number of}  & \textit{  }&  \textit{ $\checkmark$ } & \textit{ $\checkmark$ } & \textit{  } \\
        \textit{sensing SOs}   &  &  &  & \\        
        \hline
    \end{tabular}
\caption{ Experiments: $\checkmark$  - time overhead, $\checkmark$ * time and bandwidth overhead } \label{tab:expTable}
\end{table}

\subsection{Experimental results}
In executing our experiments,  we considered three main characteristics that may impact the performance of the proposed solution. These are: privacy preferences complexity, queries complexity, and number of sensing smart objects. 
However,  it was not possible to test all these dimensions in each scenario (see Table \ref{tab:expTable}). Indeed, the experiment varying the number of sensing SOs requires to use IoT-Lab dataset,  and,  as discussed in Section \ref{sec:expScenarios}, this dataset has been used only for processor SO \rom{2} and consumer SO. Experiments have been executed by varying a single dimension and keeping fixed the others with worst case settings.
Table \label{tab:expTable} presents a summary of conducted experiments. In the following, we illustrate the results.

\subsubsection{Varying the query complexity}\label{exp2} 
\begin{figure} [H]
	\centering
	\includegraphics[width=75mm]{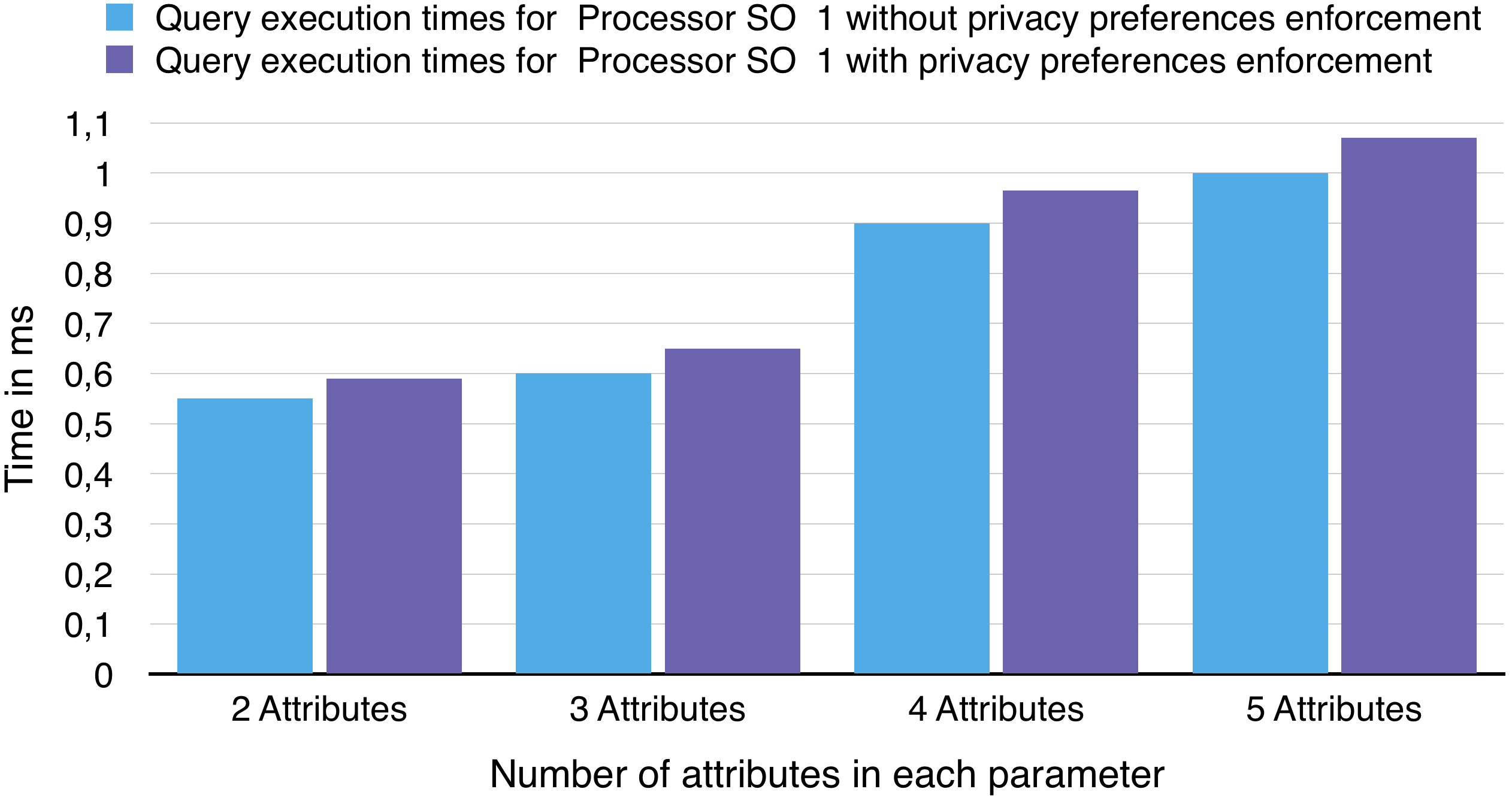}
	\caption{ Varying the query complexity - Processor SO \rom{1}} \label{fig:processor1comp}
\end{figure}

{\bf Processor SO \rom{1}:} these   SOs are only able to perform $\sigma$ and $\Pi$ operators. Moreover, since $\sigma$ operator does not alter PEAS meta-data of processed attributes (cfr. Section \ref{sec:enforcement}), we only considered the $\Pi$ operator. 
We varied  query complexity by  increasing the number of attributes to be evaluated to perform the $\Pi$ operator. We recall that, according to the adopted notation, the $\Pi$ operator is  modeled as a set of parameters, each one having a set of attributes to be evaluated. Thus, in this experiment, we have considered a $\Pi$ operator with a fixed number of parameters (i.e., 3), by varying the number of attributes in each of them.  As an example, the simplest query is defined as   a $\Pi$ operator with three parameters and two attributes for each parameter.  Given a stream S(a,b,c), an example of simplest query is $\Pi_{(a+b),(c-b),(a\times c)}$.
Execution times of queries with different complexity levels with and without the proposed  privacy preference enforcement are illustrated in Figure \ref{fig:processor1comp}. Even for the most complex queries, time overhead  is less than 0.1 ms and performance overhead is less than 7\%.

{\bf Processor SO \rom{2}:}  query complexity can be estimated by the number of  operators  it contains. However, we have also to take into account that smart objects are constrained devices, so not being  able to perform too complex SQL-like queries like those done by DBMSs. For instance, the processor SO \rom{2} has been simulated on Raspberry Pi with only 1 GB RAM. To cope with this limitation, we have considered two sets of queries, with different complexity. The first set, named {\em Simple Queries}, contains  queries consisting  only of  a $\Sigma$ operator. We select this aggregation operation as we expect it is the most common task performed by smart objects, that is,  aggregating data sensed by sensors. The second set, named {\em Complex Queries}, contains  queries  performing a join of data sensed by two clusters of sensing SOs, then a  projection over joined attributes, and, finally, an aggregation (i.e., average on 10s window) on  projected values.  Again this type of queries represents a typical operation for smart objects. Table \ref{tab:lev2tab2} shows  query execution time with and without privacy preference enforcement.  Even for the most complex queries, the time overhead is less than 0.71 ms and performance overhead is less than 9\%.

\begin{table}[H]
    \centering
    \scriptsize 
    \begin{tabular}{|c|c|c|}
        \hline
        {\bf  Processor} &{\bf  Without}  &  {\bf  Proposed}\\
        \textbf{SO \rom{2}} & {\bf PP enforcement}  &  {\bf mechanism} \\
        \hline
        \textit{Simple}  & \textit{5 ms} & \textit{5.36 ms} \\
        \textit{Queries} &   &   \\
        \hline
        \textit{Complex} & \textit{8 ms} & \textit{8.71 ms} \\
        \textit{Queries} &   &   \\
        \hline
    \end{tabular}
\caption{Varying query complexity - Processor SO \rom{2}} \label{tab:lev2tab2}
\end{table}

{\bf Consumer SO:} even if  consumer SOs are not designed to perform queries, but only compliance verification, we have to note that query complexity might impact the execution of this verification. Indeed, the more complex is the query, the more operations it contains. Thus, a  complex query  implies,   as a consequence, more complex  History fields to be evaluated by the compliance verification algorithm. As such,  the aim of this experiment is to estimate how query complexity impacts the overhead given by compliance check. For this experiment, we used the same queries generated for  processor SO \rom{2}. Results of experiments are illustrated in Table \ref{tab:lev2tab3}. Even for complex queries, the time overhead is less than 0.44 ms.

\begin{table}[H]
    \centering
    \scriptsize 
    \begin{tabular}{|c|c|c|}
        \hline
        {\bf  Scenario} &{\bf  Simple }    &  {\bf Complex }\\
          &{\bf  queries}    &  {\bf queries}\\
        \hline
        \textit{Consumer SO \rom{2}} & \textit{0.18 ms} & \textit{0.44 ms} \\
        \hline
    \end{tabular}
\caption{Varying query complexity - Consumer SO} \label{tab:lev2tab3}
\end{table}

\begin{table*}
    \centering
    \scriptsize
    \begin{tabular}{|c|c|c|c|c|c|c|c|c|c|c|}
        \hline
         \textit{ Queries} & {\bf Q1} & \textbf{Q2} &{\bf Q3}  &  {\bf Q4} & {\bf Q5} & \textbf{Q6} &{\bf Q7} & {\bf Q8}  & {\bf Q9} & \textbf{Q10} \\
        \hline
		\textit{Query selectivity} & \textit{3,5\%}	&\textit{1,84\%}&	\textit{0,85\%}&	\textit{0,72\%}&	\textit{0,55\%}&	\textit{0,37\%}&	\textit{0,28\%}&	\textit{0,2\%}	&\textit{0,15\%}&	\textit{0,13\%} \\
        \hline
     	\textit{ Extra bits per output tuple} & \textit{98}&	\textit{140}&	\textit{210}&	\textit{245}&	\textit{294}&	\textit{336}	&\textit{392}&	\textit{455}&	\textit{483}&	\textit{546} \\    
        \hline
      \textit{ Bandwidth overload per hour} &  \textit{3440}&	\textit{2576}&	\textit{1785}&	\textit{1764}&	\textit{1617}	&\textit{1243}&	\textit{1097}&	\textit{910}	&\textit{724}&	\textit{709} \\ 
        \hline 
    \end{tabular}
\caption{Varying query complexity - SO network - bandwidth overhead} \label{tab:bandwidth}
\end{table*}

\begin{figure} [H]
	\centering
	\includegraphics[width=82mm]{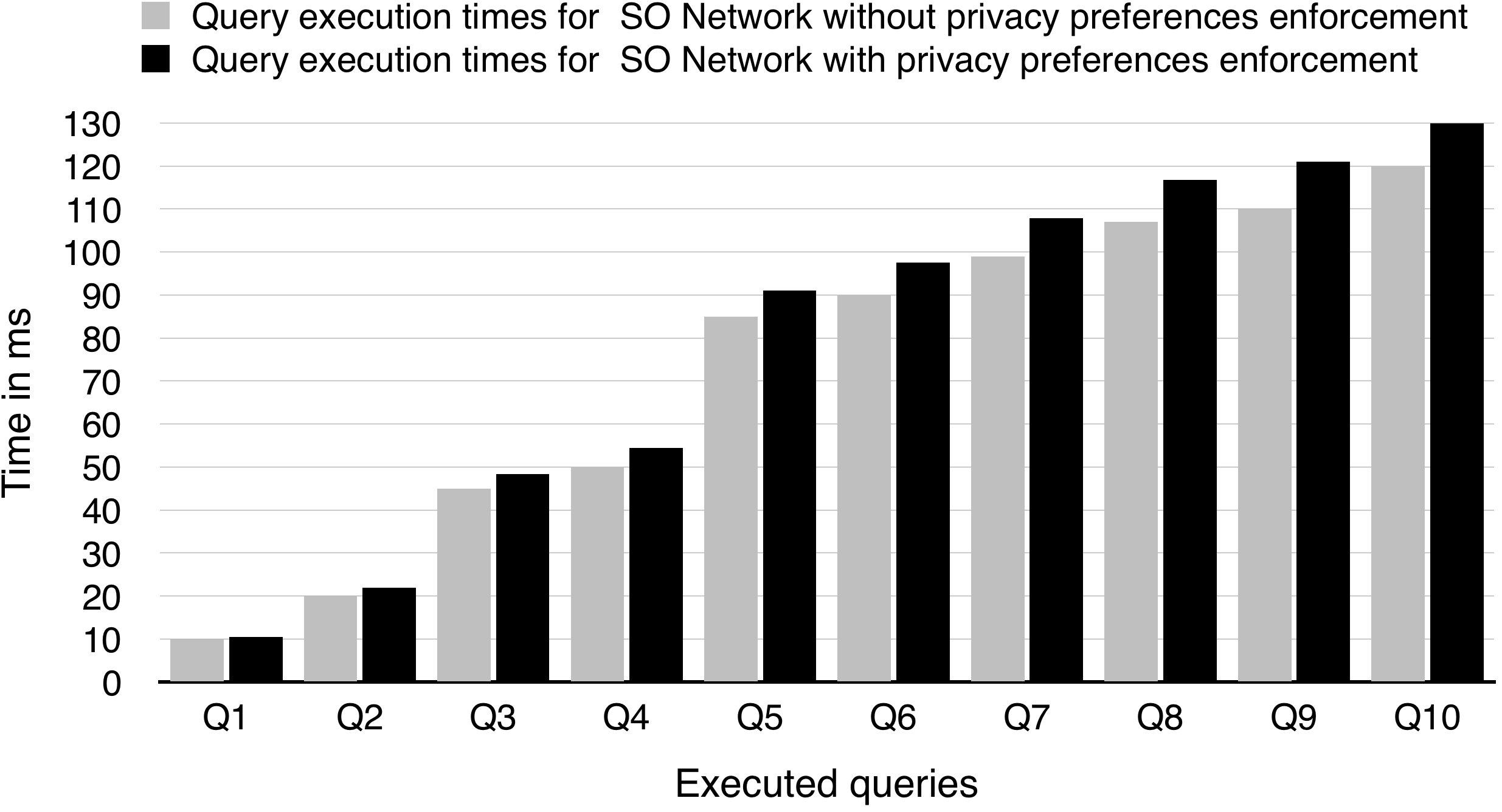}
	\caption{Varying query complexity - SO network - time overhead} \label{fig:soNetworkcomp}
\end{figure}

{\bf Smart Object network:} We simulate a smart object network via a Streambase query, where each single Streambase operator acts as a smart object. Thus,   increasing query complexity is like increasing the network complexity. In performing these experiments, we use a set of  10 queries with different numbers and types of operators. In particular, the simplest query, Q1, contains   1 $\pi$, 1 $\sigma$, 1 $\Join$,  and 1 $\Sigma$ operator, whereas the most complex query, Q10, contains  10 $\pi$, 10 $\sigma$, 10 $\Join$, and 10 $\Sigma$ operators. Results of experiments are  illustrated in Figure \ref{fig:soNetworkcomp}. As expected, less complex queries take less time to be processed. However, even in complex network scenarios, the system overhead is always  less than 10\% which is less than 12 ms.

Regarding bandwidth utilization,  we have to note that the overhead is mainly implied by meta-data inserted in the original tuples. We recall that these meta-data are generated and updated by each single operator specified in the query. As such,  we expect that a dimension impacting the bandwidth utilization is the query complexity. In addition to query complexity, another element that might impact this overhead is the query selectivity, that can be measured as the ratio of the number of tuples entering and number of tuples returned to/from a query. As query selectivity depends on  characteristics of query operators (as an example, in the aggregation operator a bigger window size implies an higher query selectivity), we have  modified the benchmark of 10 queries above described so as to tune the  operators (e.g., window size in aggregation, selection conditions in $\Pi$) to obtain an increasing selectivity.



Number of bits added into output tuple after query execution are illustrated in the second line of Table \ref{tab:bandwidth}. As expected, more complex queries increase the  number of bits added to the meta-data. However, it is relevant to note that even the most complex query consumes quite low bandwidth, 0.546 kbits, compared to the network capabilities of devices running similar queries (like Raspberry Pi).\footnote{www.pidramble.com/wiki/benchmarks/networking}

In addition to that, we want to underline that we expect that complex queries will have higher selectivity over input tuples (see first line of Table \ref{tab:bandwidth}), thus less output tuples will be generated.  Thus, on the one hand  a more complex query increases the number of added bits, but on the other hand, it decreases the number of output tuples and thus the overall bandwidth utilization. To better explain this balance, we run an experiment simulating  a smart home scenario generating 1,000 tuples per hour, each containing a float data type attribute (e.g., room temperature, used electricity in kw). This implies  32 bits in each single tuple. In case that no queries are executed, that is, all tuples are flooding the network, we have   32 kbit overall bandwidth per hour. From the third line of Table \ref{tab:bandwidth}, we can see that the overall overhead decreases by varying query complexity, aka query selectivity. Therefore, even in the worst case, output tuples generated by queries consume quite low bandwidth, i.e. 3.4 kbits.

\subsubsection{Varying complexity of privacy preferences}\label{exp1}
We recall  that a privacy preference poses a set of conditions to be verified on consumer privacy policy, namely $ip$, $jac$, and $cdc$ (see Section \ref{sec:privacyModel}). The presence of these conditions implies checks and operations both in the compliance verification algorithm as well as in the PEAS generation algorithms. As such,  we can see  the number of conditions specified in a given $pp$ as a dimension to measure  $pp$  complexity. In view of this, we generated two sets of privacy preferences with two different levels of complexity. In particular, privacy preferences in the  first set, named {\em Simple PP}, have been defined as very simple preferences  with  a unique  condition  on the intended purpose field, and thus their compliance verification is easy to be performed. In contrast,  privacy preferences in the second set, named {\em Full PP}, have been defined with  a condition on each field (i.e., $ip$, $jac$, $cdc$), so as to make harder the compliance verification.  More precisely,   privacy preferences have been defined in such a way that their constraints for  both {\em Simple PP} and {\em Full PP} refer to random elements of the data category and purpose trees, which, in turn, have been populated with synthetic data.  In particular, we have considered 100 elements in both the data category and  the  purpose trees. This test case has been applied to all experiment scenarios, as described in the following.

{\bf Processor SO  \rom{1}:} since FRDM is an hard-coded device with limited capabilities, we assumed that with each attribute in the synthetic dataset  the same privacy preference is associated (i.e., every attribute has the same $pp$ value). This  corresponds to a scenario where the implemented processor SO  \rom{1} receives as input a flow of data generated by a  sensing SO owned by a  single user, as such all data are marked with the same privacy preference. However, to test how complexity of privacy preferences impacts the local overhead of processor SO  \rom{1}, we run several times these experiments, changing in each execution the privacy preferences associated with attributes, by selecting them both from {\em Simple PP} and {\em Full PP} sets.

{\bf Processor SO \rom{2}, Consumer SO, and Smart Object network:} differently from processor SOs \rom{1},  privacy preferences are dynamically generated  and randomly assigned to  each attribute of the tuples entering in the considered smart object.

In performing these experiments  we adopt the settings shown in Table \ref{tab:case1},  which also presents the execution time of queries without privacy preferences (i.e., \textit{No PP}),  with simple privacy preferences ({\em Simple PP}), and complex privacy preferences ({\em Full PP}). As expected, simple privacy preferences  imply less time to execute the compliance verification. However, even with complex privacy preferences  the overhead is always  under 1ms and performance overhead is always less than 10\% for every scenario.  

\begin{table}
    \centering
    \scriptsize
    \begin{tabular}{|c|c|c|c|c|}
        \hline
         {\bf Scenario} & {\bf Settings} & \textbf{No} &{\bf  Simple  }    &  {\bf  Full  }\\
        &  & \textbf{PP } & {\bf    PP}    &  {\bf   PP}  \\\hline
        \textit{Processor } & Complex queries & \textit{0.31 ms} & \textit{0.325 ms} & \textit{0.34 ms} \\
        \textit{SO \rom{1}} & & & & \\
        \hline
        \textit{Processor} & Complex queries,  &\  & & \\
        \textit{SO \rom{2}} &  10 sensing SOs & \textit{3 ms} & \textit{3.10 ms} & \textit{3.28 ms} \\
        			& 100\% assoc. PP & & & \\
        \hline
       \textit{ } & Complex queries, &   &  & \\ 
        \textit{Consumer SO} & 10 sensing SOs & \textit{ $\emptyset$ } & \textit{0.08 ms} & \textit{0.34 ms} \\ 
        \textit{ } & 100\% assoc. PP &  & & \\ 
        \hline
        \textit{SO network} & Complex queries & \textit{10 ms} & \textit{10.74 ms} & \textit{10.99 ms} \\
        \hline
    \end{tabular}
\caption{Overhead by varying PP complexity} \label{tab:case1}
\end{table}

\subsubsection{Varying the number of sensing SOs}\label{exp3}




We expect that increasing the number of tuples to be processed will increase the overhead, as it increases the number of privacy preferences to be elaborated by PEAS update algorithms and the number of attributes to be checked for their compliance by consumer SO. 

\begin{figure} [H]
	\centering
	\includegraphics[width=85mm]{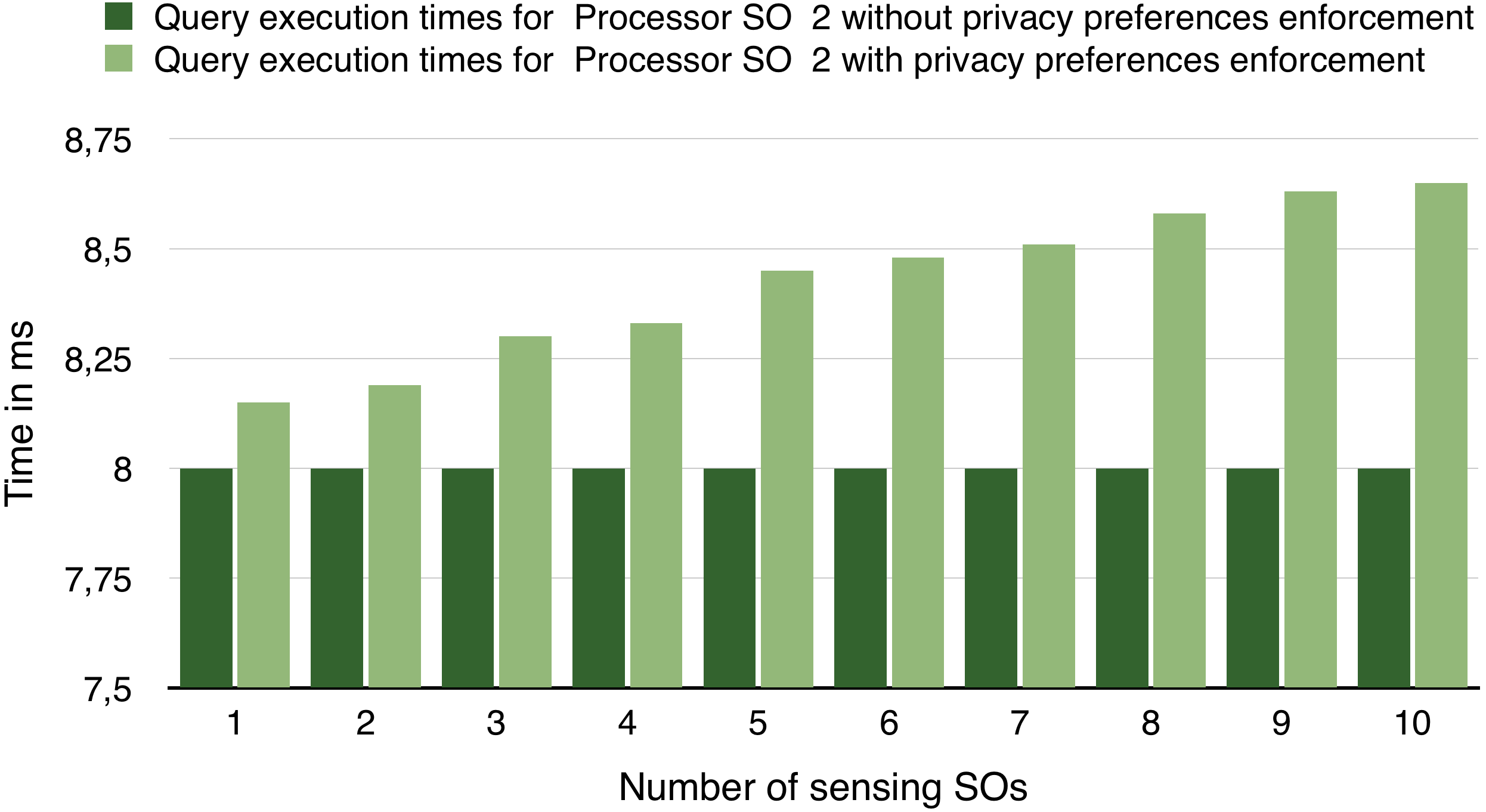}
	\caption{Varying the number of sensing SOs in Processor SO \rom{2} scenario \label{fig:numberPSO}}
\end{figure}

\begin{figure} [H]
	\centering
	\includegraphics[width=85mm]{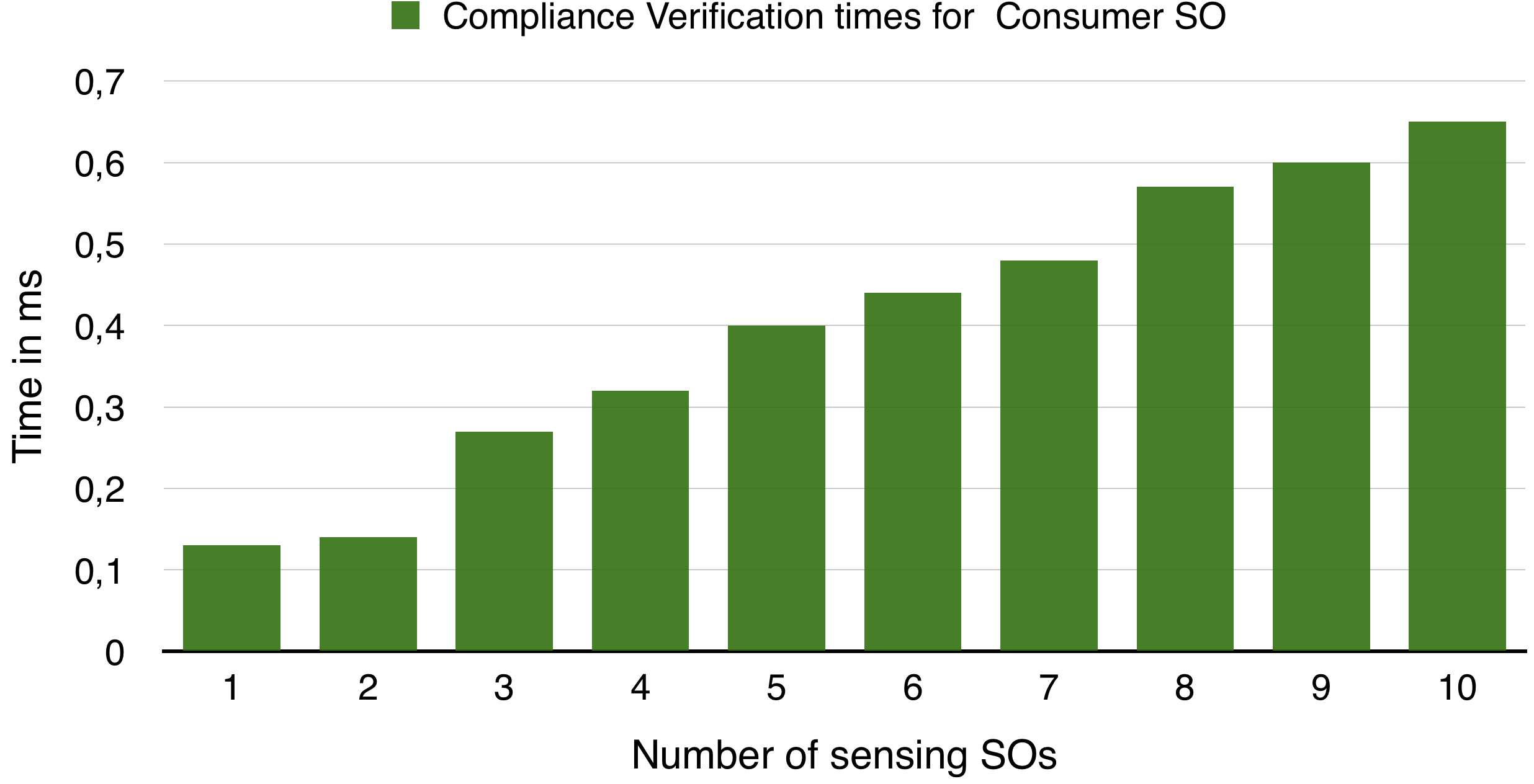}
	\caption{Varying the number of sensing SOs in Consumer SO scenario \label{fig:numberCSO}}
\end{figure}

More precisely, in both scenarios, we varied the number of nodes in IoT-Lab testbed (i.e., number of sensing SOs)  from 1 to 10, as we assume this range is reasonable for  scenarios like the smart home. In performing these experiments, we  executed the
queries in the \textit{Complex Queries} set,  described  in Section \ref{exp2}, and we associate with each sensing SO a privacy preference taken from  the {\em Full PP} set. Results of experiments are illustrated in Figure \ref{fig:numberPSO} for processor SO \rom{2}, and in Figure \ref{fig:numberCSO} for consumer SO scenarios.

As expected, for both scenarios having more nodes in the system takes more time to process. 
\footnote{We note that, we have performed similar experiments by varying the percentage of sensing SOs with an associated privacy preferences, and we have observed same overhead trends.} 
For processor SO \rom{2}, even in the worst case where we have 10 sensing SOs, overhead is under 0.7ms, thus performance overhead is less than 9\%. Compliance check performed by consumer SOs always takes less than 0.7 ms.


\section{Related Work}\label{sec:related}
In  recent years,   security and privacy in the IoT domain have been deeply investigated, with the results that various approaches have been  proposed for dealing with  different aspects of security as well as of privacy. In this section, we provide an overview of those proposals that are more related to the proposed framework. In particular, we focus on those approaches that  enforce, in some ways, users' privacy. However, we also have to note that literature offers several interesting proposals that, like our framework, deal with the problem of decentralized  policy enforcement. All these efforts have been done in domains different from IoT, but they deserve to be cited and compared to our solution. In the following, we summarize works in these two directions.

\subsection{Enforcement of users' privacy in the IoT domain}
So far different access control models have been  exploited in  the IoT domain: role based access control (RBAC) (e.g., \cite{authACIot}, \cite{rbacWoT}); capability based access control (CapBAC) (e.g., \cite{skarmeta}); attribute based access control (ABAC) (e.g., \cite{Sicari2016133}, \cite{abacEffIoT}), and access control models based on semantic rules (e.g., \cite{secoman}). Although these proposals are instrumental to control how users' personal data are used, and thus, in some sense, to protect users' privacy, they do not make users able to provide their own preferences on how their data have to be used and distributed. 
In contrast, our proposal makes user able to have a full control on how data have to be processed (e.g., accessed, aggregated, released).

User's privacy preferences have been considered in  
 \cite{torre2016framework}, which proposes a framework avoiding  inference of personal data due to  data fusion. Users specify their privacy preferences in terms of a level of confidentiality associated with each data. The proposed framework consists of a  central unit, called Personal Data Manager (PDM), that manages personal data collected by different devices,  playing thus the role of a gateway between users and third party applications. A further module, called Adaptive Interface Discovery Service (AID-S),   computes the risk of inference associated with a data disclosure, via probabilistic models and learning algorithms (e.g., RST, KNN, Bayes Filter, HMM etc.). Based on this risk value,  AID-S recommends optimal privacy settings to users to reduce the privacy risks. Similar to our proposal, also this approach considers user's perspective, but only in stating the confidentiality level of personal data, whereas our privacy model considers several dimensions of a privacy preference. Additionally, we enforce privacy of the user against data inferences in a decentralized setting, being able to pose more limitations on possible  data fusions.

Compliance  of user's privacy preferences with third party's privacy policies have been considered  in \cite{broenink2010}. Here, it has been  proposed an application for mobile phones that supports customers in making privacy decisions. Privacy preferences are  automatically generated according to the result of a questionnaire filled by users. 
The application informs the user whether his/her privacy preferences complies with the corporate's privacy policies.  In contrast, we handle privacy  in a bigger application scenario, that is, we  enforce user privacy preferences in a decentralized IoT scenario, where different smart objects may apply their own queries  over data and other parties may get involved in data processing.

Similar to our approach, other proposals have targeted smart environments (e.g., smart home and smart city systems) with the aim of protecting users' privacy.
In \cite{networkLvLSecSmartHome}, authors address the security and privacy problems of IoT smart home at the network level, that is, by monitoring network activities of IoT devices to detect suspicious behaviors. 
An external entity, called Security Management Provider (SMP) has been proposed. SMP can add access control rules to protect specific IoT devices or can apply dynamic policies to change access control rules depending on the context (e.g., the family members being present or absent from the house). 
This proposals aims at protecting privacy of the user by limiting access on data through  an external entity, i.e., SMP,  with the use of context information. In contrast, in our approach, we enforce user-defined privacy preferences to protect users' privacy in a decentralized scenario.

In \cite{ppSmartCity}, a two layered architecture is proposed for protecting users' privacy in smart city applications. A trusted layer is designed to store
real identities of individuals that can be processed only by the platform's components, without disclosing the identities to the outside world. In contrast, an untrusted second layer only makes generic, unidentifiable and identity-independent information available to external applications.
Even if this proposal  protects personal data, this is enforced only inside the trusted layer, without considering future operations that may be done on the released data  to infer new sensitive information. Moreover, users are not able to set and enforce their own privacy preferences.

\subsection{Decentralized policy enforcement}
A notable example of decentralized privacy management is represented by the sticky policy approach \cite{stickyPolicies}. 
According to  this approach user privacy preferences are strictly associated (sticky) with users' data. \cite{stickyPolicies} describes the core mechanisms required for managing sticky policies, along with Public Key Infrastructure (PKI) and encryption methodologies to attach sticky policies with data as well as to enforce them. \cite{distStickyFlexTrust} presents a distributed enforcement approach for sticky policies that permits data to be disseminated across heterogeneous hardware and software environments without pre-existing trust relationships. Also,  \cite{stickySO} presents a sticky policy approach to manage the access to IoT resources by allowing users to set and manage access control policies on their own data. In this approach, sticky policies allow to define: owner of the data; purposes for which the data can be used; a timestamp that points out the validity; and constraints which represent the rules for filtering the data with obligations and restrictions.

Our solution has some similarities with the sticky policy approach,  as we share the same goal, that is, decentralized enforcement of user privacy preferences. However,  in our proposal we go beyond the traditional  privacy preference model, where constrains are posed mainly on  purpose, retention time,  and third party usage by proposing a mechanism to derive privacy preferences for newly generated data.

Additionally, sticky policy approaches use encryption mechanisms to enhance privacy, whereas in our approach encryption is just used to secure communication. Indeed, encryption mechanisms  add extra level of complexity and demand higher resources from the devices.
However, we have to note that  cryptographic solutions have been often used to enforce distributed  access control in several domains, like  online social networks (e.g., \cite{ carminatiOSN, encryptionSN}), or cloud infrastructure (e.g., \cite{encryptionCloud,ruj2014decentralized}).
In the following we focus on efforts done in wireless sensor networks, as these are more relevant for the IoT domain.   For instance, \cite{distributedPrivacyACsensor} presents a distributed privacy preserving access control scheme designed for network of sensor nodes owned by different users, connected via a network, and managed via an offline certificate authority.
In the proposed scheme, access control is regulated exploiting tokens that users have  to pre-buy from the network owner before entering the sensor network. Users can query sensor data with unspent tokens and sensor nodes are subject to validate the token and grant appropriate amount of requested data. The  scheme makes use of blind signatures in token generation, which ensures that tokens are publicly verifiable yet they are unlikable to user identities.
Similarly to the previous cited work, \cite{distributedACPrivacy} presents a  distributed protocol for achieving privacy-preserving access control for sensor networks by exploiting a ring signature scheme. 
To query sensor nodes, a user needs to build a ring signature along with the query command and send them together to sensor nodes. 
However such approaches do not consider  privacy preferences of the data owner.

Since one of the features of the proposed framework is to derive  privacy preferences for new  data, we have to note that 
literature offers also works dealing with composition of heterogeneous access control policies. These have been done with the goal of composing a set of access control policies into a single one \cite{semanticAlgebraAC, algebraComposingAC}.
For instance, \cite{semanticAlgebraAC} proposes a semantic framework for policy composition without  committing to a specific access control model. Access control policies are modelled as four valued predicates. Similarly, \cite{algebraComposingAC} proposes an algebra for constructing  security policy from different policies.  

\vspace{-5pt}
\section{Conclusion}\label{sec:conclusion}
In this paper, we proposed a decentralized  privacy enforcement mechanism for IoT, where compliance check of user individual privacy preferences is performed directly by smart objects. The experimental evaluation  has shown a negligible time and bandwidth overhead. 

We plan to extend our proposal according to several directions. 
We recall that the proposed  decentralized  privacy enforcement mechanism assumes the existence of a set of  "derivation paths", that is, a set of inference rules able to predict the data category resulting from the execution of an operator. As a first future work, we aim at investigating how to derive these rules. 
At this purpose, we plan to exploit existing data fusion schemes proposed for IoT domain \cite{fusionAlgoIoT, ahmad2016efficient}, where different methodologies  have been designed   (e.g., probability based, artificial intelligence based) aiming at combining  data from multiple sensors to produce more accurate, more complete, and more dependable information that could not be possible to achieve through a single sensor. 
These methods could be deployed in our framework to model the data flow, and thus derivation paths. 
Specifically,  multi-dimensional data fusion algorithms, such as \cite{ahmad2016efficient,fusionAlgoIoT}, could be used to  extract association rules from data generated by subnet of sensing SOs.
Second, we plan to implement a hybrid approach for compliance check, that combines apriori and aposteriori compliance verification, to decrease the performance overhead of the system.

We also plan to improve usability of the system. Indeed, we are aware that an average user of an IoT application may not be skilled enough to understand potential privacy threats and to properly set up his/her  privacy preferences. At this purpose, we plan to investigate tools to help users in setting their privacy preferences, based on learning from their habits, as it has been previously done in other domains (e.g.,  Online Social Networks \cite{privacyOSN}).

\bibliographystyle{model1-num-names}
\bibliography{biblio}

\end{document}